\documentclass[preprint,12pt,authoryear]{elsarticle}

\usepackage[utf8]{inputenc}
\DeclareUnicodeCharacter{00D7}{\ensuremath{\times}}
\usepackage[T1]{fontenc}
\usepackage{amsmath,amssymb,bm}
\usepackage{graphicx}
\usepackage{booktabs}
\usepackage{tabularx}
\usepackage{array}
\usepackage{multirow}
\usepackage[hidelinks]{hyperref}
\usepackage{enumitem}
\usepackage{microtype}
\usepackage{subcaption}
\usepackage[table]{xcolor}
\usepackage{adjustbox}
\usepackage{flafter}

\definecolor{rowblue}{RGB}{243,240,235}

\graphicspath{{figures/}}
\makeatletter
\def\ps@pprintTitle{%
  \let\@oddhead\@empty
  \let\@evenhead\@empty
  \let\@oddfoot\@empty
  \let\@evenfoot\@empty}
\makeatother
\begin{document}

\begin{frontmatter}

\title{Whose readiness counts? Disagreement within and between sectors\\
in perceived AI and robotics preparedness}

\author[surrey]{Peng Wang\corref{cor1}}
\ead{peng.wang@surrey.ac.uk}
\cortext[cor1]{Corresponding author.}
\address[surrey]{Centre for Vision, Speech and Signal Processing (CVSSP), University of Surrey, Guildford GU2 7XH, United Kingdom}

\begin{abstract}

AI and Industry~4.0 readiness assessments often summarise preparedness using a single score for an organisation, application domain or sector. Those summaries can conceal disagreement about the same technology and variation among applications grouped under one sector label. We test how much information is lost through this aggregation using a card-based survey in which 982 respondents provided 15{,}200 readiness evaluations across 17 named AI and robotics challenges. Readiness is perceived community preparedness and available resources, not personal willingness or audited organisational capability. Respondents frequently disagreed about identical challenges, with card-level readiness standard deviations of $1.03$-$1.26$ on a five-point scale. A crossed decomposition attributes 32.7\% of observed variation to stable respondent differences, 7.3\% to differences among challenges, and 60.0\% to response-level variation that also contains measurement error. Differences among challenge-family means account for only about 2\% of variation, with substantially more variation among people, applications and person-family judgements. Manufacturing has the highest mean readiness, yet shop-floor robotics, process-optimisation AI and general decision-support applications are judged differently. Computer-science and AI/ML respondents report higher readiness than non-technical respondents across challenge families, whereas engineering respondents do not report higher Manufacturing readiness. Sector rankings are therefore best used as portfolio summaries rather than evidence that an industry is uniformly ready or behind. Readiness reporting should retain application-level disagreement, disclose whose judgements form the average, and consider ethics, cyber security, literacy and capability needs without collapsing them into a single score.

\end{abstract}


\end{frontmatter}

\section{Introduction}

AI and robotics are increasingly assessed not only in terms of technical performance, but also in terms of whether organisations, sectors and communities are prepared to adopt, govern and support them. Readiness assessments therefore play an important role in decisions about investment, workforce development, deployment priorities and governance. In practice, however, these assessments are often communicated through a small number of summary indicators: an organisation receives a readiness level, a factory is assigned a maturity score, or a sector is described as more or less prepared than another.

This approach is well established. Industry~4.0 maturity models summarise multiple organisational and technological conditions into overall readiness levels \citep{Schumacher2016}. Organisational AI-readiness frameworks similarly combine factors such as strategy, data, skills, culture, leadership and technical resources into broader preparedness constructs \citep{NewJohnk2021,NewHolmstrom2022,NewAliKhan2025}. Sector-specific studies have applied related ideas in areas such as exhibitions, healthcare and manufacturing \citep{NewHradecky2022,NewDuus2026,Schumacher2016}. These summaries are useful because they compress complex evidence into a form that decision makers can compare and communicate.

The limitation is that, while a summary score can describe the centre of a set of judgements, it can hide the structure underneath it. The same mean readiness score can arise from very different situations. Respondents may largely agree that preparedness is moderate, or the same mean may combine people who see strong preparedness with others who see a serious shortfall. A similar problem arises when several applications are grouped under one sector label. Battery quality control and catalyst-process optimisation may both be classified as Manufacturing, while a robot chef, humanoid housekeeper and smart travel assistant may all be classified as Hospitality, yet the technologies, implementation conditions and perceived risks are clearly different. A sector average is most informative when the applications and respondents behind it are judged similarly; it becomes less informative when substantial disagreement exists within the sector label.

Existing research provides strong reasons to expect such heterogeneity. Technology acceptance and readiness research shows that individuals differ in their willingness or propensity to adopt and use new technologies \citep{Davis1989,Venkatesh2003,Venkatesh2012,Parasuraman2000}. AI-specific studies further show that attitudes and trust depend on the application, organisational role and context being considered rather than on `AI' as a single object \citep{NewHorowitz2024,NewDaly2025,NewNikolova2025}. Hospitality research similarly shows that applications sharing a sector label can involve different service roles, operational implications and adoption challenges \citep{Ivanov2017,Tuomi2021}. Technology assessment and responsible innovation research therefore argues for keeping plural appraisals visible rather than reducing them prematurely to a single framing \citep{Stirling2008,Stilgoe2013}. What remains less clear is how much of the variation in a readiness exercise actually lies between broad challenge families, how much lies between applications within those families, and how much reflects differences among the people making the judgements.

This distinction matters for policy because sector rankings can easily be interpreted as evidence that one industry is more prepared than another. If differences among broad group means account for a large share of the observed variation, such rankings may provide a reasonable summary. If they account for only a small share, however, then a league table is not simply a compressed version of the evidence: it omits much of what determines the observed preparedness judgements. The same issue applies within organisations. A high readiness score may partly reflect who was asked, which applications were included and whether specialist and non-specialist groups interpreted community capability in the same way. Readiness assessment therefore needs to establish not only \emph{how high the score is}, but also \emph{how much agreement surrounds it, what is being averaged, and whose judgements form the average}.

We address this problem using a repeated card-based survey in which 982 participants provided 15{,}200 evaluations of 17 named AI and robotics challenges. The challenges span three application domains, i.e., Biomedical, Manufacturing and Hospitality, together with Cyber Security and Ethical Awareness. The latter two are analysed as cross-cutting considerations rather than treated as additional sectors. Each challenge was rated on five-point measures of significance, complexity and readiness. The present paper focuses on readiness, defined as perceived community preparedness and available resources rather than personal willingness or independently audited organisational capability. Because many respondents evaluated the same challenges and multiple applications within the same challenge family, the design allows us to examine disagreement at the challenge, application, family and respondent levels rather than relying only on aggregate means. The work therefore addresses four research questions:

\begin{enumerate}[leftmargin=1.6em,itemsep=2pt]

\item How much do respondents disagree when evaluating the same AI or robotics challenge?

\item How much readiness variation lies between challenge families, compared with variation among applications and respondents within them?

\item After accounting for general respondent and card effects, how are the application domains related to Ethical Awareness and Cyber Security, and should the latter be treated as one preparedness object or as distinct considerations?

\item To what extent do professional backgrounds shift readiness assessments, and do these differences vary across challenge families?

\end{enumerate}

The contributions of this work are threefold. First, it moves beyond evaluating readiness using a single sector or domain score by decomposing where readiness variation actually occurs. This provides direct evidence on whether differences among challenge families capture most of the structure in the data or only a small part of it. Second, it tests whether applications grouped under the same sector label are judged sufficiently similarly to support a single preparedness summary, with particular attention to Biomedical, Manufacturing and Hospitality. Third, it examines whose judgements shape those summaries by comparing professional backgrounds and by separating application-domain patterns from the cross-cutting roles of Ethical Awareness and Cyber Security.

These contributions are intended to make readiness evidence more useful for decision making. For policymakers, the results indicate when sector-level scores are suitable for broad portfolio scanning and when funding, skills or regulatory decisions require more disaggregated evidence. For sector bodies, they show why a high average should not automatically be presented as sector-wide consensus. For organisations, they highlight the importance of reporting which applications were assessed and whose views contributed to the score. A readiness measure is therefore most informative when it answers not only \emph{which sector appears ready}, but also \emph{ready for which application, according to whom, and with how much agreement}.

This paper is a further step from a companion analysis of the same survey \citep{Paper1TFSC}, which examines why the same person rates different challenges differently. The present paper asks how judgements vary among people and among applications within and between sectors.

\section{Data and methods}

\subsection{Survey design and measures}

We analyse a cleaned card-based survey of perceived AI and robotics preparedness: 982 participants and 15{,}200 usable evaluations after exclusion based on attention checks and missing data; 864 participants completed all 17 challenges. Participants were recruited through Prolific between 12 and 29 June 2026. Eligibility was adult Prolific member. Of 1{,}235 registered users, 240 provided no challenge response, leaving 995 respondents with challenge data. The platform attention-check indicator excluded 13 of those respondents, yielding the 982-person primary sample. Analyses requiring a balanced person-by-challenge matrix use the 864 complete-deck respondents and are labelled accordingly. The study was not preregistered. Participants provided informed consent on the platform and were compensated according to the Prolific policy; ethics for the survey were approved by the relevant committee at the author's institution.

Each card presents a concrete AI or robotics challenge together with three core five-point ratings, which appeared on each card and were not independently randomised. The survey cards and measures are as follows:

\begin{itemize}[leftmargin=1.4em,itemsep=2pt]

\item Significance ($S$): potential impact of addressing the challenge (1 = very low impact; 5 = critical impact).

\item Complexity ($C$): combined technical, organisational and societal difficulty (1 = very simple; 5 = very complex), considering barriers, coordination, societal factors and resource requirements.

\item Readiness ($R$): current community preparedness and available resources (1 = very poor readiness; 5 = excellent readiness), considering initiatives, educational resources, research progress and community engagement.

\end{itemize}

Readiness is therefore not personal willingness, competence or independently audited organisational capability. The survey did not define a common geographic or institutional boundary for `community'. The reason is twofold: it is difficult to define a single boundary that applies across all challenges, and the contextual information on each challenge card was expected to guide respondents' judgements. Optional comments were stored as free text and were not compulsory. Missing challenge responses are handled by pairwise deletion for correlations ($n=864$-$945$ per pair) and by using available cards within a challenge family for person-family means (4{,}532 observed person-family records of 4{,}910 possible). Complete-deck analyses drop respondents who did not rate all 17 challenges.

The challenges are grouped into five challenge families. In the analysis we distinguish three application domains, i.e., Biomedical, Manufacturing and Hospitality, from Ethical Awareness and Cyber Security, which are retained as cross-cutting areas because they should be considered when application decisions are made. Professional background and other demographics are recorded at participant level (Table~\ref{tab:sample}).

The primary outcome for person $i$ and challenge $j$ is $R_{ij}$, representing that person's readiness rating for the challenge. Between-respondent dispersion for a challenge is summarised by the full five-category distribution, mean, standard deviation (SD) and interquartile range (IQR). We define the person-level challenge-family mean as
\[
\bar R_{if}=\frac{1}{J_f}\sum_{j\in f}R_{ij},
\]
\noindent
which averages that person's available ratings within challenge family $f$. The standard deviation of the same person's ratings inside a multi-card family describes observed application-level heterogeneity within that family. It is not interpreted as internal-consistency reliability or proof that a family is, or is not, a latent construct. Ethical Awareness is treated as one observed indicator rather than an ethics scale. The three Cyber Security cards are summarised as a composite only after checking their empirical association.

\begin{table}[htbp]
\centering
\caption{Participant profile for the analytical sample ($N=982$).}
\label{tab:sample}
\scriptsize
\renewcommand{\arraystretch}{0.95}
\setlength{\tabcolsep}{3pt}

\begin{minipage}[t]{0.27\textwidth}
\textbf{Gender}\\[2pt]
\rowcolors{2}{white}{rowblue}
\begin{tabularx}{\linewidth}{@{}Xrr@{}}
\toprule
 & $n$ & \% \\
\midrule
Female & 494 & 50.3 \\
Male & 477 & 48.6 \\
Non-binary & 8 & 0.8 \\
Prefer not to say & 3 & 0.3 \\
\bottomrule
\end{tabularx}
\rowcolors{2}{white}{white}

\vspace{8pt}

\textbf{Age}\\[2pt]
\rowcolors{2}{white}{rowblue}
\begin{tabularx}{\linewidth}{@{}Xrr@{}}
\toprule
 & $n$ & \% \\
\midrule
18-24 & 134 & 13.6 \\
25-34 & 360 & 36.7 \\
35-44 & 272 & 27.7 \\
45-54 & 123 & 12.5 \\
55-64 & 76 & 7.7 \\
65+ & 17 & 1.7 \\
\bottomrule
\end{tabularx}
\rowcolors{2}{white}{white}
\end{minipage}
\hfill
\begin{minipage}[t]{0.38\textwidth}
\textbf{Professional background}\\[2pt]
\rowcolors{2}{white}{rowblue}
\begin{tabularx}{\linewidth}{@{}Xrr@{}}
\toprule
 & $n$ & \% \\
\midrule
Non-technical & 288 & 29.3 \\
Business & 163 & 16.6 \\
Other & 117 & 11.9 \\
Research & 76 & 7.7 \\
Operations & 73 & 7.4 \\
Engineering & 53 & 5.4 \\
Computer science & 49 & 5.0 \\
Data science & 48 & 4.9 \\
Software development & 46 & 4.7 \\
AI/ML & 38 & 3.9 \\
Design & 31 & 3.2 \\
\bottomrule
\end{tabularx}
\rowcolors{2}{white}{white}
\end{minipage}
\hfill
\begin{minipage}[t]{0.29\textwidth}
\textbf{Education}\\[2pt]
\rowcolors{2}{white}{rowblue}
\begin{tabularx}{\linewidth}{@{}Xrr@{}}
\toprule
 & $n$ & \% \\
\midrule
Bachelor's & 415 & 42.3 \\
Master's & 233 & 23.7 \\
Some college & 160 & 16.3 \\
High school & 91 & 9.3 \\
Other & 83 & 8.5 \\
\bottomrule
\end{tabularx}
\rowcolors{2}{white}{white}

\vspace{8pt}

\textbf{Industry}\\[2pt]
\rowcolors{2}{white}{rowblue}
\begin{tabularx}{\linewidth}{@{}Xrr@{}}
\toprule
 & $n$ & \% \\
\midrule
Technology & 136 & 13.8 \\
Other & 130 & 13.2 \\
Education & 105 & 10.7 \\
Healthcare & 101 & 10.3 \\
Retail & 100 & 10.2 \\
Remaining & 410 & 41.8 \\
\bottomrule
\end{tabularx}
\rowcolors{2}{white}{white}

\vspace{8pt}

\textbf{Experience}\\[2pt]
\rowcolors{2}{white}{rowblue}
\begin{tabularx}{\linewidth}{@{}Xrr@{}}
\toprule
 & $n$ & \% \\
\midrule
Reported & 728 & 74.1 \\
Missing & 254 & 25.9 \\
\bottomrule
\end{tabularx}
\rowcolors{2}{white}{white}
\end{minipage}

\renewcommand{\arraystretch}{1.0}
\end{table}

\subsection{Analytical approach}

The analysis follows the four research questions and moves from disagreement about the same challenge, to the distribution of variation within and between challenge families, and then to cross-domain and professional-background patterns.

\subsubsection{Same-challenge disagreement}

For each of the 17 challenges, we summarise the full distribution of readiness ratings using the mean, standard deviation, interquartile range and normalised Shannon entropy. Entropy is divided by $\log 5$ so that values lie on a common scale and captures distributional spread without assuming equal distances between the five response categories. To show how average readiness and disagreement coexist, challenges are also placed on a mean-by-dispersion map. Median splits of the 17 challenge means and SDs are used only as descriptive reference lines and do not define population thresholds, clusters or absolute consensus. As a missingness check, challenge-level SDs are recomputed for the 864 participants who completed all 17 cards.

\subsubsection{Within- and between-family variation}

Application-level heterogeneity within each multi-card challenge family is summarised by the mean, across respondents, of the respondent-specific within-family SD. Because families contain different numbers and types of applications, this is interpreted descriptively rather than as a psychometric test of family coherence. As a complementary measure that is less sensitive to the number of cards, we also calculate the mean pairwise absolute difference among applications within each family. A sensitivity analysis that excludes Foundation AI Decision Support was carried out to check if sector-neutral application would change the results, as its Manufacturing assignment is the most ambiguous.

Variation in readiness is then partitioned in two complementary ways. First, a crossed null model is estimated on the complete-deck subset:
\begin{equation}
R_{ij}=\mu+u_i+v_j+\varepsilon_{ij},
\label{eq:crossed}
\end{equation}
\noindent
where $u_i$ represents stable between-respondent variation, $v_j$ represents differences among the 17 observed challenges, and $\varepsilon_{ij}$ is the remaining response-level component. The model is estimated using a balanced generalizability theory method of moments \citep{Brennan2001}. Because each respondent rates each challenge only once, the residual cannot separate person-by-challenge interaction from measurement error and is therefore not interpreted as a pure interaction component. Percentile 95\% confidence intervals are obtained from 400 respondent-level bootstrap resamples and are conditional on the 17 challenges included in the survey.

In addition, we use an approximate additive challenge-family decomposition:
\begin{equation}
R_{ifj}
=
\mu
+
F_f
+
P_i
+
C_{j(f)}
+
(P\times F)_{if}
+
\varepsilon_{ifj},
\label{eq:nested}
\end{equation}
\noindent
where $F_f$ represents differences among challenge-family means, $P_i$ represents stable differences in respondents' overall readiness ratings, $C_{j(f)}$ represents differences among applications within families, $(P\times F)_{if}$ captures how a respondent's rating of a particular family differs from their general rating tendency, and $\varepsilon_{ifj}$ represents the remaining card-level variation within respondent-family combinations. This decomposition provides the main within- versus between-family summary. A one-way analysis of variance on the full sample checks the respondent share, and a crossed mixed model fitted to the full sample provides an additional check on the complete-deck decomposition.

\subsubsection{Cross-domain structure and the roles of Ethics and Cyber Security}

We next examine how readiness assessments co-vary across the three application domains and the two cross-cutting areas, Ethical Awareness and Cyber Security. Among the 864 complete-deck respondents, observed correlations are first calculated among respondent-level challenge-family means. To distinguish substantive cross-domain alignment from a respondent's general tendency to give higher or lower ratings, we then fit the additive model
\[
R_{ij}=\mu+a_j+g_i+e_{ij},
\]
\noindent
where $a_j$ captures stable card differences and $g_i$ captures each respondent's general rating tendency. Residuals $e_{ij}$ are averaged within each challenge family for each respondent, and correlations are then calculated among these respondent-level residual profiles. Confidence intervals are obtained from 400 respondent-level bootstrap resamples.

To examine whether Ethical Awareness and Cyber Security provide distinct information about application-domain readiness, we fit a grouped binomial generalised linear model using the complete-deck respondents. Let $i$ index respondents and $d$ index the three application domains (Biomedical, Manufacturing and Hospitality). $E_i$ is respondent $i$'s Ethical Awareness readiness rating, which is based on a single observed card. $K_i$ is the mean of that respondent's three Cyber Security readiness ratings, used after confirming that the three cards are sufficiently associated to support a composite summary.

For each respondent-domain pair, $Y_{id}$ is the number of application cards in domain $d$ rated 4 or 5, out of $J_d$ cards in that domain. We model
\[
Y_{id}\sim\mathrm{Binomial}(J_d,p_{id}),
\]
with
\[
\begin{aligned}
\mathrm{logit}(p_{id})=
\alpha
+\beta_E E_i
+\beta_K K_i
+\beta_{EK}E_iK_i
+\mathrm{domain}_d
+\mathrm{domain}_d\text{ interactions}.
\end{aligned}
\]
\noindent
Here $p_{id}$ is the modelled probability that an application card in domain $d$ receives a high readiness rating ($R\geq4$) from respondent $i$. $\beta_E$ and $\beta_K$ represent the associations of Ethical Awareness and Cyber Security preparedness with high application readiness, while $\beta_{EK}$ tests whether their joint association departs from additivity on the log-odds scale. Domain terms allow the baseline probability and these associations to differ across Biomedical, Manufacturing and Hospitality. Standard errors are clustered by respondent because each respondent contributes observations for multiple application domains. All estimates are interpreted associationally rather than causally.

\subsubsection{Professional-background differences}

Professional background patterns are examined using descriptive background-by-challenge profiles and a prespecified background-by-family model. Professional backgrounds are grouped as non-technical, engineering, computer science, AI/ML, research, design, business and other, with the remaining reported backgrounds combined into the `other' category.

The inferential model uses person-family mean readiness, with one observation for respondent $i$ in each observed challenge family $f$. Let $B_{ik}$ denote an indicator that respondent $i$ belongs to professional-background group $k$, and let $F_{if}$ denote the challenge-family indicators for respondent $i$'s observation in family $f$. The model is then in the form of 
\[
\bar R_{if}
=
\alpha
+\sum_{k=1}^{7}\beta_k B_{ik}
+\sum_{f=1}^{4}\gamma_f F_{if}
+\sum_{k=1}^{7}\sum_{f=1}^{4}\delta_{kf}B_{ik}F_{if}
+\varepsilon_{if},
\]
\noindent
where $\bar R_{if}$ is respondent $i$'s mean readiness rating for challenge family $f$. Non-technical background is the reference professional group and Manufacturing is the reference challenge family. Accordingly, $\alpha$ is the expected mean readiness for non-technical respondents in Manufacturing; $\beta_k$ represents the difference between professional-background group $k$ and non-technical respondents within Manufacturing; $\gamma_f$ represents the difference between challenge family $f$ and Manufacturing among non-technical respondents; and $\delta_{kf}$ indicates how the background contrast for group $k$ changes in challenge family $f$ relative to Manufacturing.

Standard errors are clustered by respondent because each respondent contributes observations for multiple challenge families. A joint Wald test of
\[
H_0:\delta_{kf}=0 \quad \text{for all } k,f
\]
assesses whether professional-background contrasts vary across challenge families. Family-specific background contrasts are then derived from the fitted model and accompanied by Benjamini-Hochberg false-discovery-rate adjusted $p$-values across the full contrast grid. These comparisons are associational and adjusted for challenge family only. Failure to reject the joint interaction test is interpreted as absence of evidence that background contrasts differ across families, rather than evidence that the contrasts are equivalent.

\subsubsection{Exploratory network and free-text analyses}

Two exploratory analyses provide additional descriptive context. First, an undirected network of the 17 challenges is constructed from between-person Spearman correlations in readiness. An edge is displayed when $\rho\geq0.40$, an analyst-selected threshold used for visualisation rather than inference, and greedy modularity is used to summarise the resulting graph. Because these correlations can partly reflect respondents' general rating tendencies, the network is not interpreted as a recovered latent taxonomy. The stability of the Ethics-Cyber grouping is examined across thresholds from $0.30$ to $0.50$ and across 100 respondent-level bootstrap samples at the $0.40$ threshold.

Second, optional free-text comments are tagged using a coarse keyword lexicon covering ethics and fairness, privacy and security, trust, oversight, training, jobs, human experience and efficiency. Theme shares are descriptive among observed comments because participation is self-selected, tags may overlap, and the lexicon is not a validated qualitative coding instrument. The analysis is therefore used only as supporting context for interpreting the quantitative patterns.

\section{Results}

The results that follow are not a ranking of sectors or challenge families. They show how much variation such summaries can conceal.

\subsection{Same-challenge dispersion}

\begin{figure}[htbp]
\centering
\includegraphics[width=0.95\textwidth,height=0.62\textheight,keepaspectratio]{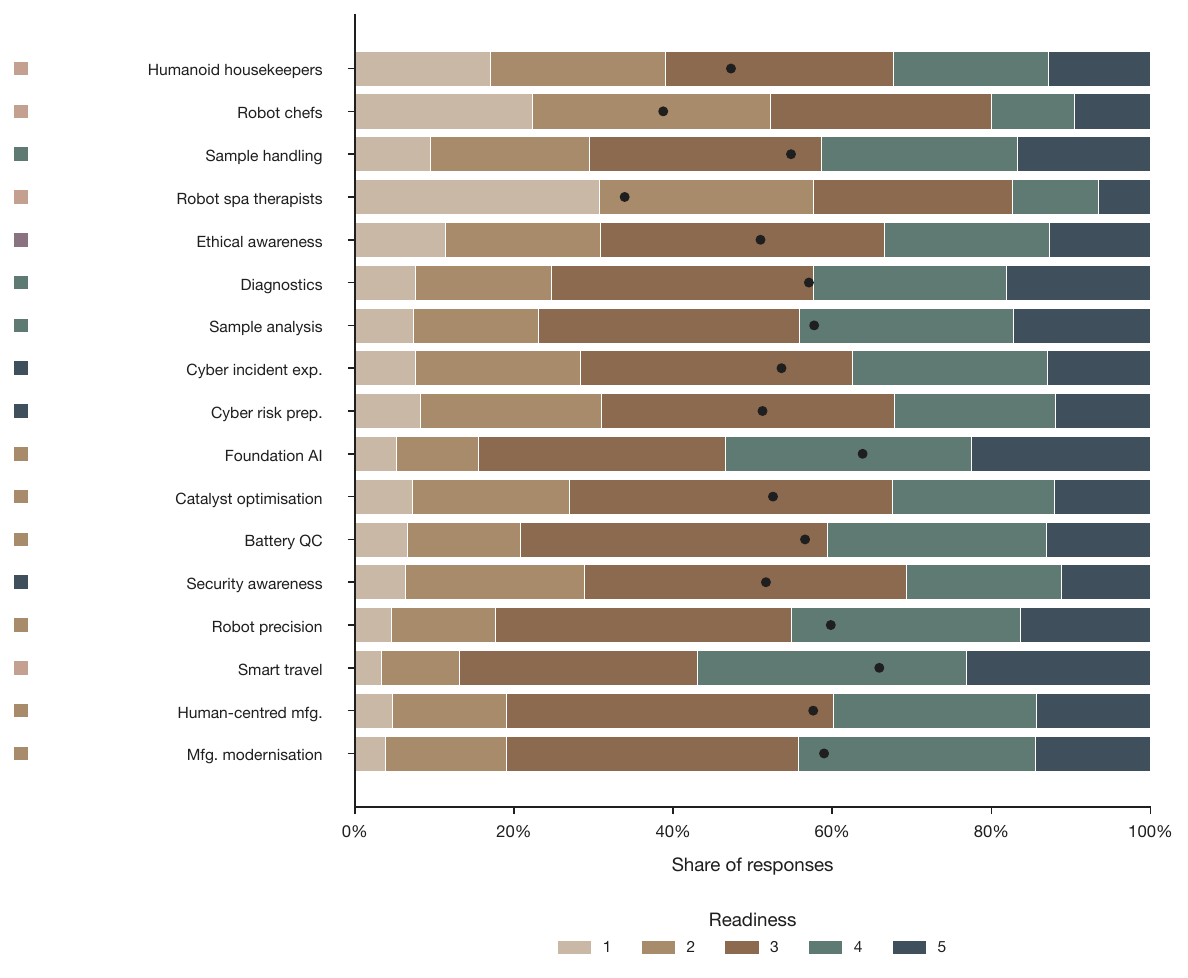}
\caption{Same-card readiness distributions (1-5). Bars show response shares; black markers indicate card means mapped onto the unit interval. Cards are ordered by readiness SD (greater dispersion toward the top). Family colour ticks mark challenge family.}
\label{fig:likert}
\end{figure}

Figure~\ref{fig:likert} shows readiness distributions for all 17 cards. Table~\ref{tab:disp} reports the mean, standard deviation and interquartile range. Readiness SDs range from approximately 1.03 to 1.26. The greatest dispersion occurs for Humanoid housekeepers, Robot chefs and Sample handling; Manufacturing modernisation, Human-centred manufacturing and Smart travel are comparatively less dispersed. Normalised Shannon entropy of the same 1-5 distributions gives a similar ordering (range $0.87$-$0.98$). Complete-deck SDs ($n=864$ respondents) reproduce the full-sample ordering almost exactly (Spearman $\rho=0.99$).

\begin{table}[htbp]
\centering
\caption{Card-level readiness mean, SD and IQR, with exploratory mean$\times$dispersion labels (median splits).}
\label{tab:disp}
\scriptsize
\begin{adjustbox}{max width=\textwidth}
\rowcolors{2}{white}{rowblue}
\begin{tabular}{llrrrrl}
\toprule
Family & Challenge & $n$ & Mean & SD & IQR & Mean$\times$dispersion \\
\midrule
Hospitality & Humanoid housekeepers & 907 & 2.89 & 1.26 & 2.00 & dispersed lower \\
Hospitality & Robot chefs & 905 & 2.55 & 1.22 & 1.00 & dispersed lower \\
Biomedical & Sample handling & 867 & 3.19 & 1.21 & 2.00 & dispersed higher \\
Hospitality & Robot spa therapists & 898 & 2.36 & 1.21 & 2.00 & dispersed lower \\
Ethics & Ethical awareness & 888 & 3.04 & 1.17 & 2.00 & dispersed lower \\
Biomedical & Diagnostics & 870 & 3.28 & 1.17 & 1.00 & dispersed higher \\
Biomedical & Sample analysis & 867 & 3.31 & 1.15 & 1.00 & dispersed higher \\
Cyber Security & Cyber incident experience & 918 & 3.14 & 1.12 & 2.00 & dispersed lower \\
Cyber Security & Cyber risk preparedness & 949 & 3.05 & 1.11 & 2.00 & dispersed lower \\
Manufacturing & Foundation AI support & 872 & 3.55 & 1.10 & 1.00 & shared higher \\
Manufacturing & Catalyst optimisation & 881 & 3.10 & 1.08 & 2.00 & shared lower \\
Manufacturing & Battery QC & 877 & 3.26 & 1.07 & 1.00 & shared higher \\
Cyber Security & Security awareness & 976 & 3.07 & 1.06 & 2.00 & shared lower \\
Manufacturing & Robot precision & 868 & 3.39 & 1.05 & 1.00 & shared higher \\
Hospitality & Smart travel & 911 & 3.64 & 1.04 & 1.00 & shared higher \\
Manufacturing & Human-centred manufacturing & 874 & 3.30 & 1.03 & 1.00 & shared higher \\
Manufacturing & Manufacturing modernisation & 872 & 3.36 & 1.03 & 1.00 & shared higher \\
\bottomrule
\end{tabular}
\rowcolors{2}{white}{white}

\end{adjustbox}
\end{table}

\begin{figure}[htbp]
\centering
\includegraphics[width=0.88\textwidth,height=0.42\textheight,keepaspectratio]{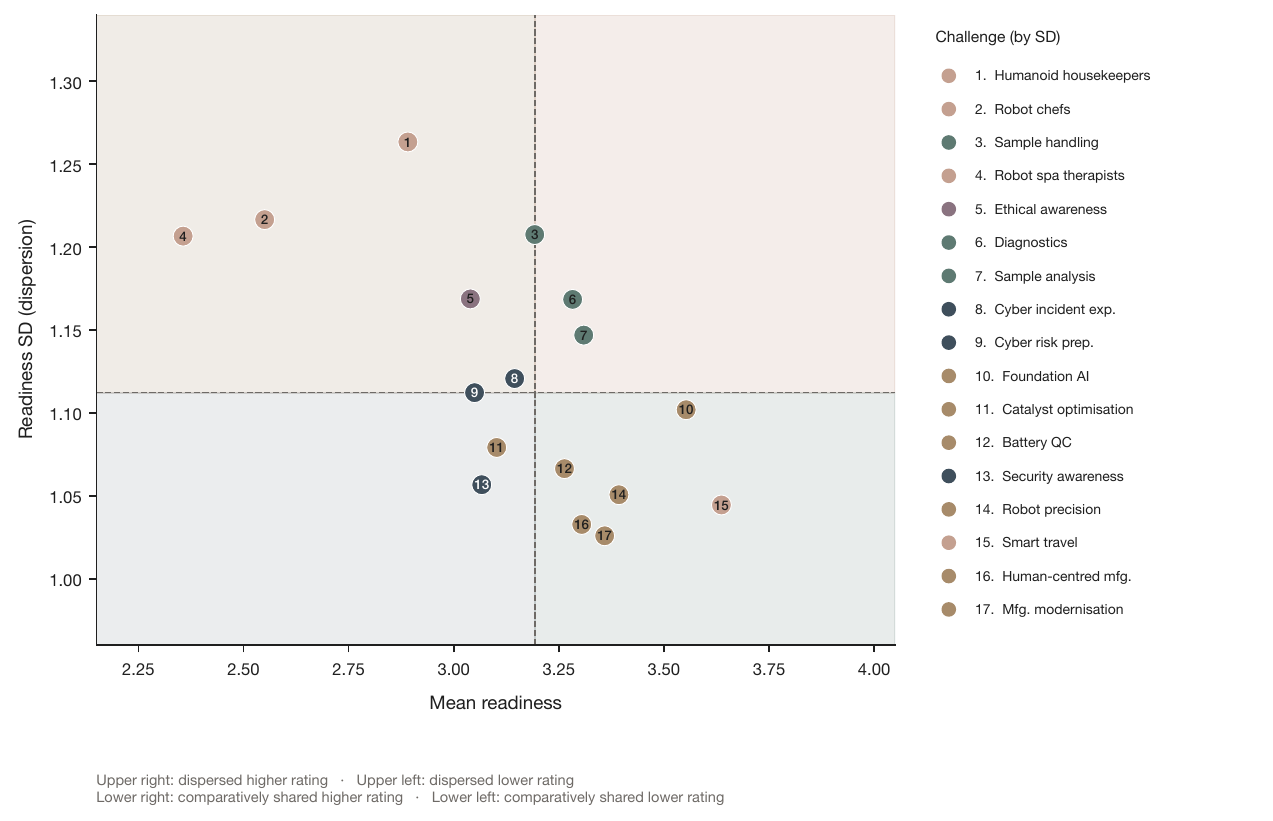}
\caption{Mean$\times$dispersion map of card-level readiness. Points are numbered by decreasing readiness SD (greatest dispersion $=1$). Dashed lines mark the medians of the 17 means and SDs. Quadrant labels are relative exploratory descriptions, not consensus thresholds or confirmatory clusters. Marker colours follow challenge family.}
\label{fig:meanmap}
\end{figure}

Figure~\ref{fig:meanmap} places each card in the mean$\times$dispersion plane. Several Hospitality cards fall in the lower-mean/higher-dispersion quadrant, whereas several Manufacturing cards fall in the higher-mean/lower-dispersion quadrant. The Manufacturing exception is catalyst-process optimisation, which is the only Manufacturing card in the shared-lower cell. Smart travel is an informative Hospitality exception, combining higher mean readiness with comparatively lower dispersion. Because the reference lines are sample medians, these labels are relative descriptions rather than absolute evidence of consensus.

\subsection{Sources of variation}

\begin{figure}[htbp]
\centering
\begin{subfigure}[t]{0.54\textwidth}
\centering
\includegraphics[width=\linewidth,height=0.28\textheight,keepaspectratio]{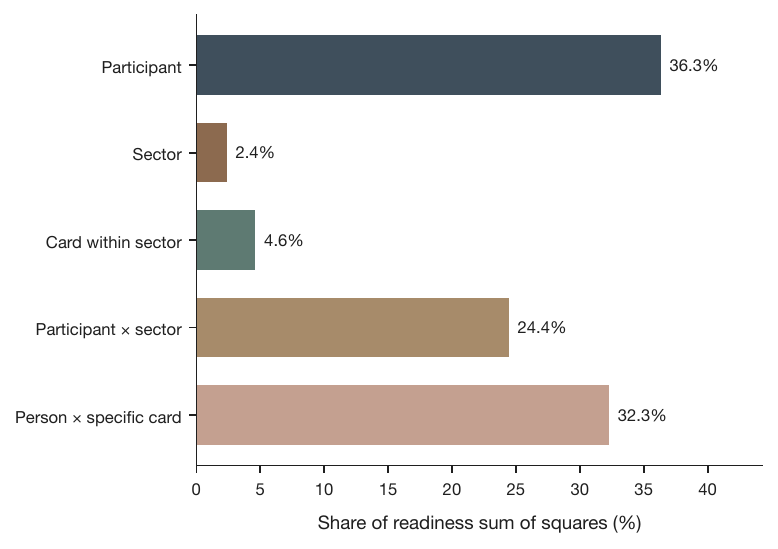}
\caption{Challenge-family partition (approx.)}
\label{fig:var:ss}
\end{subfigure}\hfill
\begin{subfigure}[t]{0.44\textwidth}
\centering
\includegraphics[width=\linewidth,height=0.28\textheight,keepaspectratio]{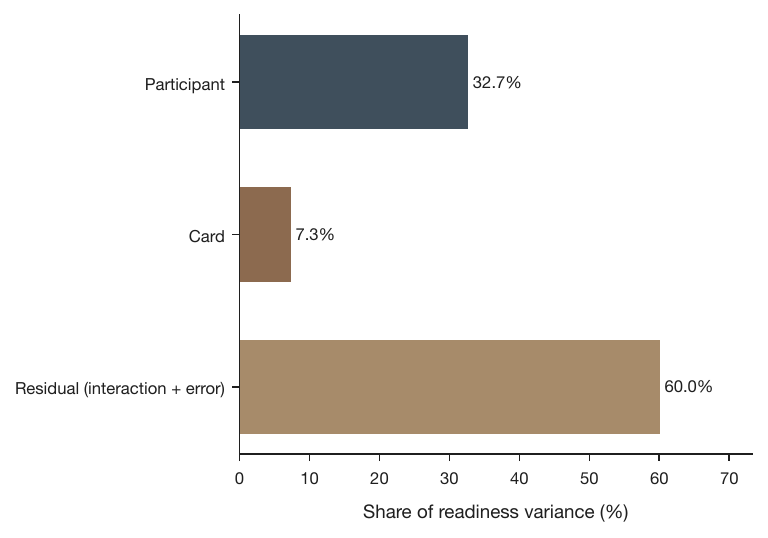}
\caption{Crossed person$\times$challenge}
\label{fig:var:gt}
\end{subfigure}
\caption{Sources of variation in perceived readiness on the complete-deck subset. (a)~Approximate challenge-family sum-of-squares partition: family, application within family, person$\times$family, and leftover residual. (b)~Crossed respondent-by-challenge decomposition. The residual in (b) combines person-by-challenge interaction and measurement error because each cell contains one rating.}
\label{fig:var}
\end{figure}

Figure~\ref{fig:var} and Table~\ref{tab:var} summarise both partitions. Conditional on the complete-deck respondents and the 17 observed challenges, the crossed decomposition assigns 32.7\% of readiness variance to stable respondent differences (person-bootstrap 95\% CI 29.9-35.6), 7.3\% to differences among challenges (6.4-8.3), and 60.0\% (57.4-62.6) to response-level residual variation. Because there is no replicate rating within a person-challenge cell, this residual combines person-by-challenge interaction, transient judgement and measurement error; it should not be interpreted as a pure interaction component. A one-way analysis of variance on the full sample yields a respondent share of 0.325. A crossed mixed model on the full sample gives essentially the same respondent/challenge/residual split (33\%/7\%/60\%).

The family-level partition answers the within- versus between-family question. Differences among challenge-family means are small ($\approx 2.4\%$), and differences among applications within a family are also modest ($\approx 4.6\%$). Much more of the observed variation reflects how a person treats a challenge family ($\approx 24.4\%$) and leftover card-specific residual variation ($\approx 32.3\%$). In other words, differences among broad family means are a small part of the story. What matters more is who is judging, how they judge a family as a whole, and how they respond to particular applications within it. The nested shares are an additive approximation without bootstrap intervals; they are read alongside the crossed decomposition rather than as a competing causal model.

\begin{table}[htbp]
\centering
\caption{Variance decompositions for readiness on the complete-deck subset. Nested block: approximate challenge-family sum-of-squares partition. Crossed block: person-by-challenge decomposition with person-level bootstrap 95\% intervals ($B=400$). Residuals include measurement error.}
\label{tab:var}
\small
\rowcolors{2}{white}{rowblue}
\begin{tabular}{llrr}
\toprule
Model & Component & Share (\%) & 95\% CI \\
\midrule
\multicolumn{4}{l}{\emph{A. Nested partition (complete deck; approximate)}} \\
 & Participant & 36.3 & - \\
 & Sector (between) & 2.4 & - \\
 & Application within sector & 4.6 & - \\
 & Participant $\times$ sector & 24.4 & - \\
 & Residual (card leftover $+$ error) & 32.3 & - \\
\midrule
\multicolumn{4}{l}{\emph{B. Crossed person$\times$challenge (complete deck; bootstrap)}} \\
 & Participant & 32.7 & [29.9, 35.6] \\
 & Challenge & 7.3 & [6.4, 8.3] \\
 & Residual (interaction $+$ error) & 60.0 & [57.4, 62.6] \\
\bottomrule
\end{tabular}
\rowcolors{2}{white}{white}

\end{table}

\subsection{Within- and between-family structure}

\begin{figure}[htbp]
\centering
\includegraphics[width=0.58\textwidth,height=0.32\textheight,keepaspectratio]{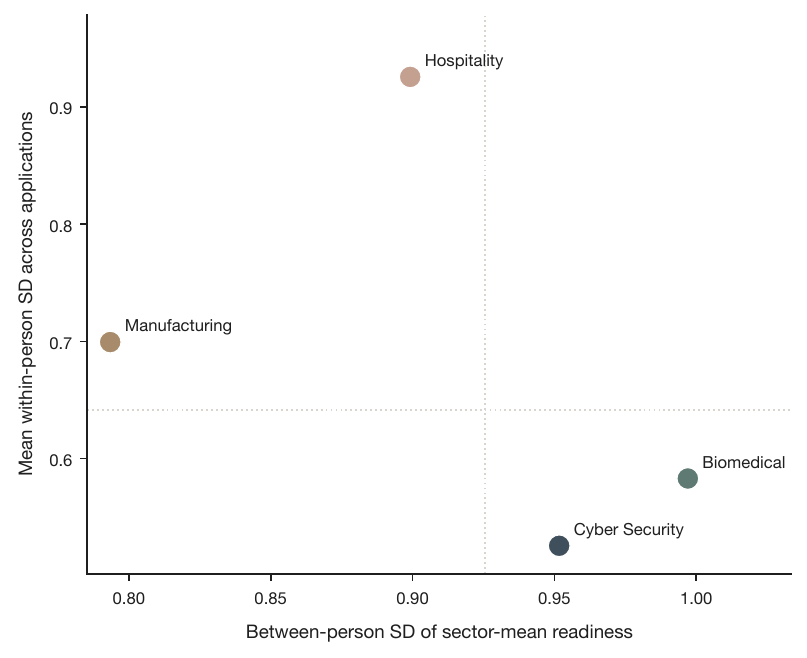}
\caption{Between-person SD of person-level challenge-family mean readiness versus mean within-person SD across applications in that family (Ethical Awareness omitted; single card). Comparisons are descriptive because families contain different applications and numbers of cards.}
\label{fig:sector}
\end{figure}

Figure~\ref{fig:sector} and Table~\ref{tab:sector} make the family-level result concrete. They contrast between-person dispersion in challenge-family means with within-person variation across applications inside each multi-card family. Hospitality has the largest observed mean within-person SD, approximately $0.93$. In this instrument, respondents therefore vary their ratings more across housekeeping, food-service, spa and travel applications than across the applications grouped in the other multi-card families. This pattern is consistent with hospitality-robotics research showing application-specific judgements \citep{Ivanov2017,Tuomi2021}.

\begin{table}[htbp]
\centering
\caption{Challenge-family structure: between-person dispersion versus within-person application variation.}
\label{tab:sector}
\small
\begin{adjustbox}{max width=\textwidth}
\rowcolors{2}{white}{rowblue}
\begin{tabular}{lrrrr}
\toprule
Sector & Cards & Mean $R$ & Between-person SD & Mean within-person application SD \\
\midrule
Manufacturing & 6 & 3.33 & 0.79 & 0.70 \\
Biomedical & 3 & 3.26 & 1.00 & 0.58 \\
Hospitality & 4 & 2.86 & 0.90 & 0.93 \\
Cyber Security & 3 & 3.09 & 0.95 & 0.53 \\
Ethics & 1 & 3.04 & 1.17 & - \\
\bottomrule
\end{tabular}
\rowcolors{2}{white}{white}

\end{adjustbox}
\end{table}

\begin{figure}[htbp]
\centering
\begin{subfigure}[t]{0.48\textwidth}
\centering
\includegraphics[width=\linewidth,height=0.34\textheight,keepaspectratio]{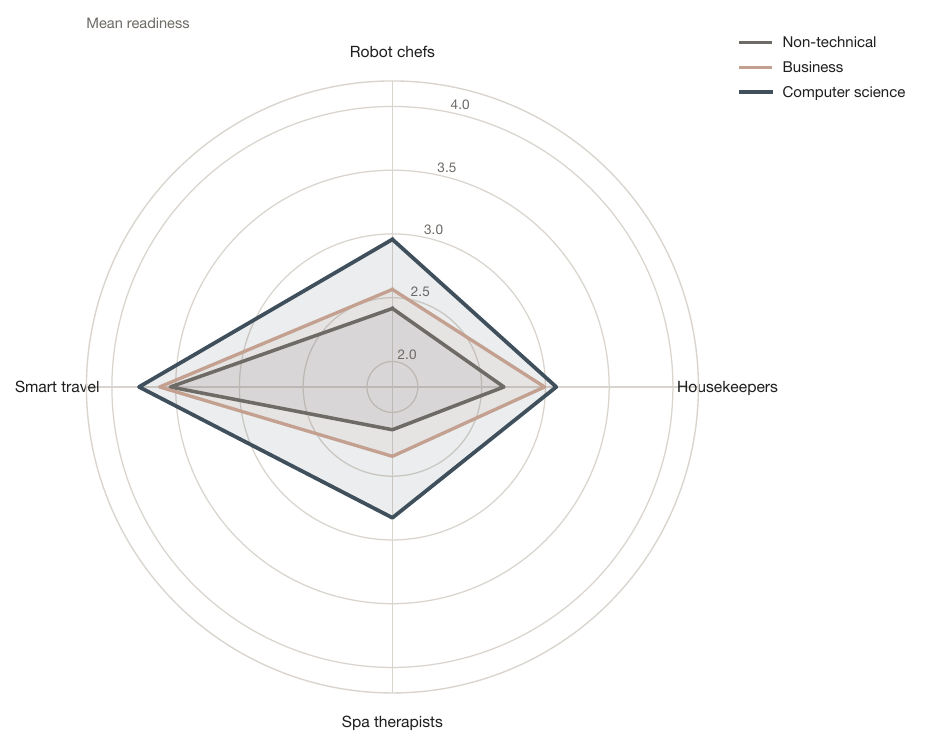}
\caption{Hospitality}
\label{fig:hospradar}
\end{subfigure}\hfill
\begin{subfigure}[t]{0.48\textwidth}
\centering
\includegraphics[width=\linewidth,height=0.34\textheight,keepaspectratio]{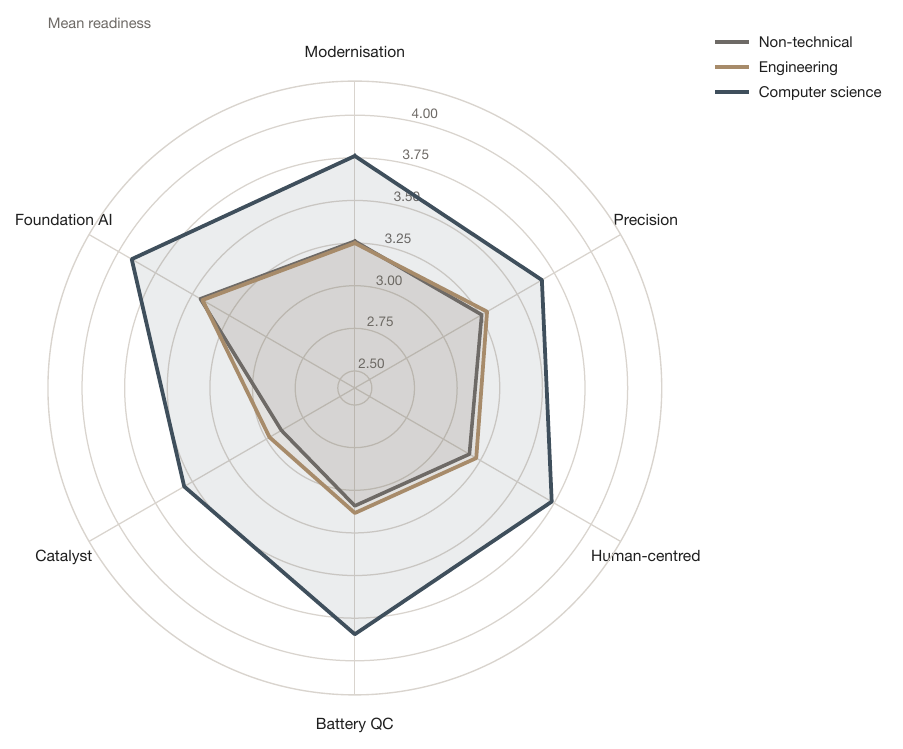}
\caption{Manufacturing}
\label{fig:mfgradar}
\end{subfigure}
\caption{Application-domain profiles rather than single scores. Axes are the challenges inside each application domain; polygons are unadjusted mean readiness for selected professional backgrounds. (a)~Smart travel sits apart from the three hospitality service-robot cards. (b)~Engineering tracks the non-technical profile on every manufacturing axis; computer science sits outside both. Catalyst-process optimisation is the low vertex; Foundation AI is the high vertex and is only loosely tied to the robotics cards.}
\label{fig:profiles}
\end{figure}

Figure~\ref{fig:hospradar} shows the same within-family split as a profile: Smart travel sits well above the three robotics-service cards for non-technical, business and computer-science respondents alike.

Manufacturing is the other pole of the same comparison, and it is the more important pole for industrial robotics. It has the highest family mean (approximately $3.33$) and a lower mean within-person SD (approximately $0.70$) than Hospitality. That combination can look like a licence to report one manufacturing readiness number. The card-level evidence does not support that reading. Every Manufacturing challenge still has a between-person SD above $1.0$. Within the family, three layers are visible (Figure~\ref{fig:mfgradar}; Table~\ref{tab:disp}). Shop floor and industrial robotics cards, such as manufacturing modernisation, robot precision, human-centred manufacturing and battery quality control, form the more coherent core: among respondents who rated all four, the mean within-person SD is $0.61$, and 40\%-45\% of ratings on modernisation and precision are 4 or 5. Catalyst-process optimisation is the Manufacturing exception in the opposite direction from Smart travel: it is the only Manufacturing card in the shared-lower quadrant (mean $3.10$), and only 32.5\% of its ratings are 4 or 5. Foundation AI Decision Support is the highest-mean card in the family (mean $3.55$; 53.4\% rated 4 or 5) and the most weakly correlated with the robotics cards (Spearman $\rho=0.34$-$0.42$). Excluding it reduces the Manufacturing mean within-person SD from 0.70 to 0.65. In other words, a high manufacturing average mixes a relatively shared industrial-robotics core, a lagging process-optimisation application, and a sector-neutral foundation model that is only loosely tied to the shop floor.

Cyber Security and Biomedical have lower observed within-person variation than Hospitality. Ethical Awareness contains one challenge, so within-family dispersion is undefined. A mean pairwise absolute difference, which is less sensitive to the number of cards than a within-family SD, yields the same family ordering: Hospitality $1.07$, Manufacturing $0.76$, Biomedical $0.70$, Cyber Security $0.64$ (Supplementary Table~S2). Restricting Manufacturing to the four shop floor cards lowers that pairwise gap to $0.70$. These comparisons describe the sampled applications; they do not establish that a whole family is a unidimensional or multidimensional construct.

\begin{figure}[htbp]
\centering
\includegraphics[width=0.58\textwidth,height=0.34\textheight,keepaspectratio]{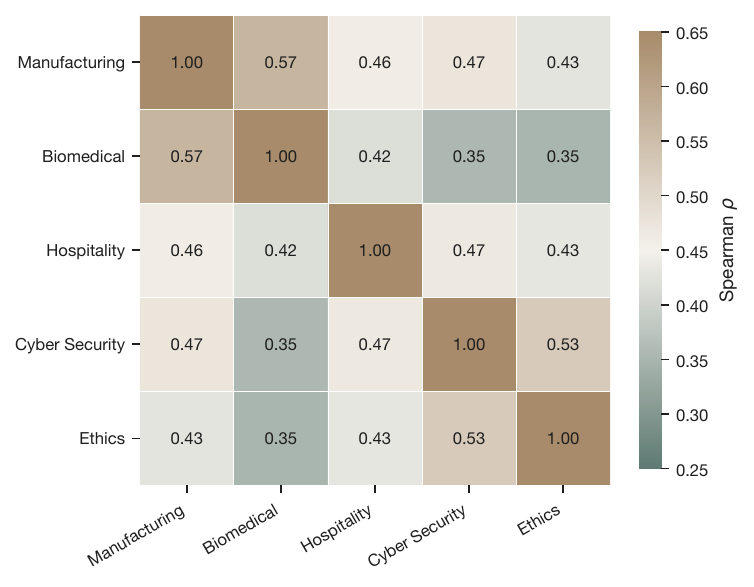}
\caption{Spearman correlations among person-level challenge-family mean readiness scores.}
\label{fig:scorr}
\end{figure}

Person-level correlations among challenge-family mean profiles (Figure~\ref{fig:scorr}; Table~\ref{tab:scorr}) are positive and moderate. The strongest association is Biomedical-Manufacturing ($\rho\approx 0.57$); Biomedical-Cyber Security is weaker ($\rho\approx 0.35$). Manufacturing therefore travels with the other industrial application domain more than Hospitality does. These correlations show that respondents who rate one family higher often rate others higher as well. They do not, by themselves, establish either a single general readiness trait or distinct latent family constructs.

\begin{table}[htbp]
\centering
\caption{Spearman correlations among person-level challenge-family mean readiness scores.}
\label{tab:scorr}
\small
\rowcolors{2}{white}{rowblue}
\begin{tabular}{llr}
\toprule
Sector A & Sector B & Spearman $\rho$ \\
\midrule
Manufacturing & Biomedical & 0.57 \\
Cyber Security & Ethics & 0.53 \\
Manufacturing & Cyber Security & 0.47 \\
Hospitality & Cyber Security & 0.47 \\
Manufacturing & Hospitality & 0.46 \\
Hospitality & Ethics & 0.43 \\
Manufacturing & Ethics & 0.43 \\
Biomedical & Hospitality & 0.42 \\
Biomedical & Cyber Security & 0.35 \\
Biomedical & Ethics & 0.35 \\
\bottomrule
\end{tabular}
\rowcolors{2}{white}{white}

\end{table}

\subsection{Observed and respondent-adjusted cross-domain connections}

\begin{figure}[htbp]
\centering
\includegraphics[width=0.96\textwidth,height=0.40\textheight,keepaspectratio]{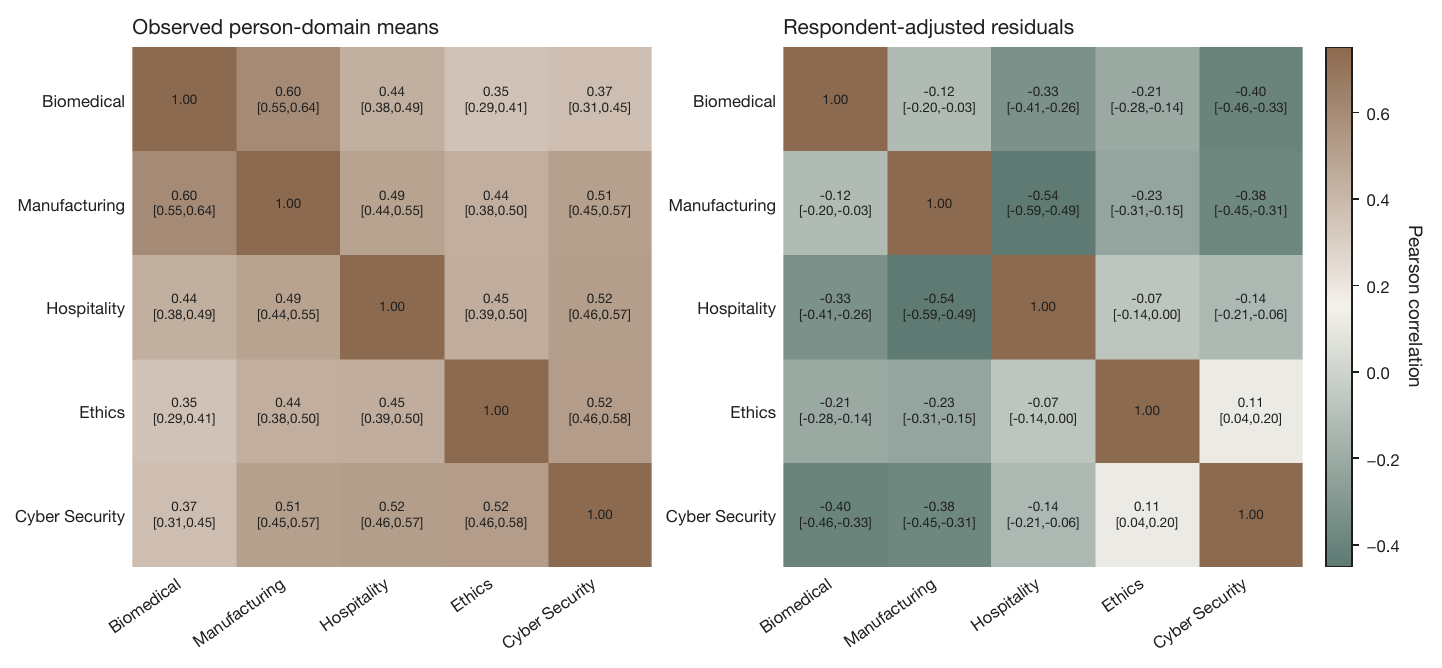}
\caption{Observed versus respondent-adjusted cross-domain readiness connections among the 864 complete-deck respondents. (a)~Left: Pearson correlations among respondent-level challenge-family means. (b)~Right: correlations among residual family profiles after a joint respondent-and-card additive model removes card differences and each respondent's general rating tendency. Cell entries show $r$ and person-bootstrap 95\% intervals.}
\label{fig:crossdomain}
\end{figure}

Figure~\ref{fig:crossdomain} compares the raw domain structure with the respondent-adjusted residual structure among the 864 complete-deck respondents. In the left panel, the observed challenge-family mean correlations are all positive. Biomedical and Manufacturing are most strongly connected ($r=0.60$, bootstrap 95\% CI $0.55$-$0.64$), and Ethical Awareness and Cyber Security are also positively associated ($r=0.52$, $0.46$-$0.58$). This is the structure that would be seen in an ordinary domain-score dashboard.

In the right panel, after accounting for card differences and each respondent's general tendency to give higher or lower ratings, the pattern changes. Most application-domain residual correlations are negative, which means that above-average residual readiness in one domain tends to accompany below-average residual readiness in another once the respondent's general level has been removed. The clearest remaining positive residual connection is between Ethical Awareness and Cyber Security ($r=0.11$, $0.04$-$0.20$). Hospitality's residual links with Ethics are near zero ($r=-0.07$, $-0.14$ to $0.00$), and its links with Biomedical and Manufacturing are negative. These adjusted results do not prove that ethics and cyber security causally underpin application readiness, but they support treating the raw Ethics-Cyber association as more than a simple general high-rating tendency.

\subsection{Ethics and Cyber: considered in decisions, not one object}

\begin{figure}[htbp]
\centering
\includegraphics[width=0.96\textwidth,height=0.40\textheight,keepaspectratio]{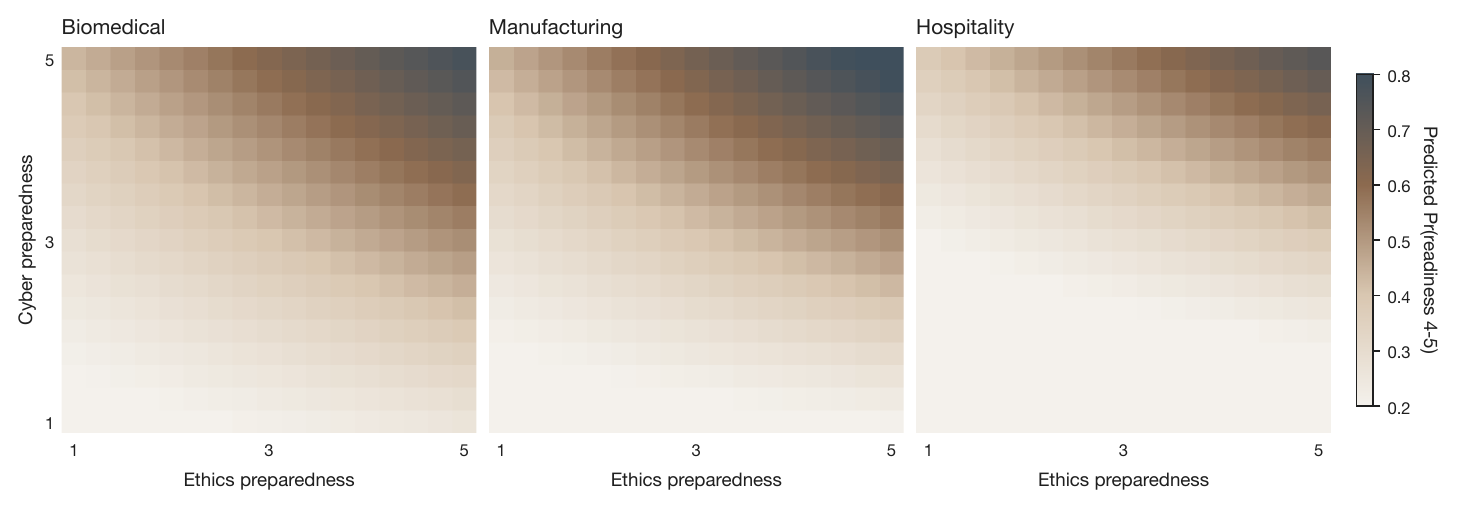}
\caption{Ethical Awareness and Cyber Security preparedness as predictors of high application readiness. Panels show predicted probabilities of a readiness rating of 4 or 5 in Biomedical, Manufacturing and Hospitality, using complete-deck respondents. Cyber Security is the mean of three Cyber cards; Ethical Awareness is one observed card.}
\label{fig:ecpred}
\end{figure}

The three Cyber Security cards support a composite summary in the complete-deck subset: Cronbach's $\alpha=0.83$, with pairwise Pearson correlations of $0.61$-$0.64$. Ethical Awareness remains a single observed indicator. Using these measures, Figure~\ref{fig:ecpred} shows predicted probabilities that respondents rate application-domain readiness as high ($R\geq4$).

The heatmaps show a broadly additive pattern. Higher Ethical Awareness preparedness is associated with higher application readiness ($p<.001$), and higher Cyber Security preparedness is also associated with higher application readiness ($p<.001$). The Ethics$\times$Cyber interaction is not clearly supported ($p=0.27$), and there is no evidence that the interaction differs across Biomedical, Manufacturing and Hospitality ($p=0.85$). Predicted high-readiness probabilities are lowest in the lower-left corner of each panel and highest in the upper-right corner. For example, at Ethics $=3$ and Cyber $=3$, predicted high-readiness probabilities are about $0.40$ for Biomedical, $0.39$ for Manufacturing and $0.28$ for Hospitality; at Ethics $=5$ and Cyber $=5$, they rise to approximately $0.77$, $0.82$ and $0.73$, respectively. These are associations between perceived ethics and cyber preparedness and perceived application readiness, not evidence that improving ethics or cyber readiness would cause application readiness to rise.

\subsection{Professional background}

\begin{figure}[htbp]
\centering
\includegraphics[width=0.62\textwidth,height=0.36\textheight,keepaspectratio]{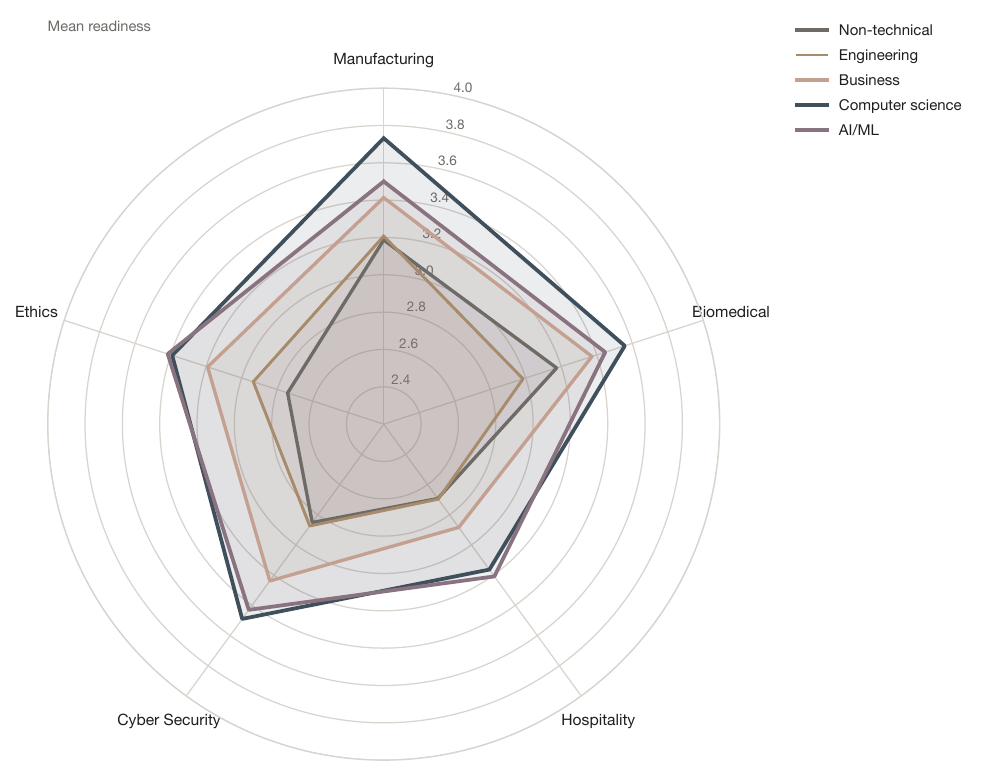}
\caption{Unadjusted mean readiness by professional background across the five challenge families (radar profile; selected groups). Similar shape with different size indicates a general gradient rather than a family-specific rearrangement. Non-technical respondents form the inner polygon: more cautious community-readiness appraisals on every axis. Computer-science and AI/ML sit outside; engineering tracks the non-technical profile.}
\label{fig:bgsec}
\end{figure}

\begin{figure}[htbp]
\centering
\includegraphics[width=0.96\textwidth,height=0.32\textheight,keepaspectratio]{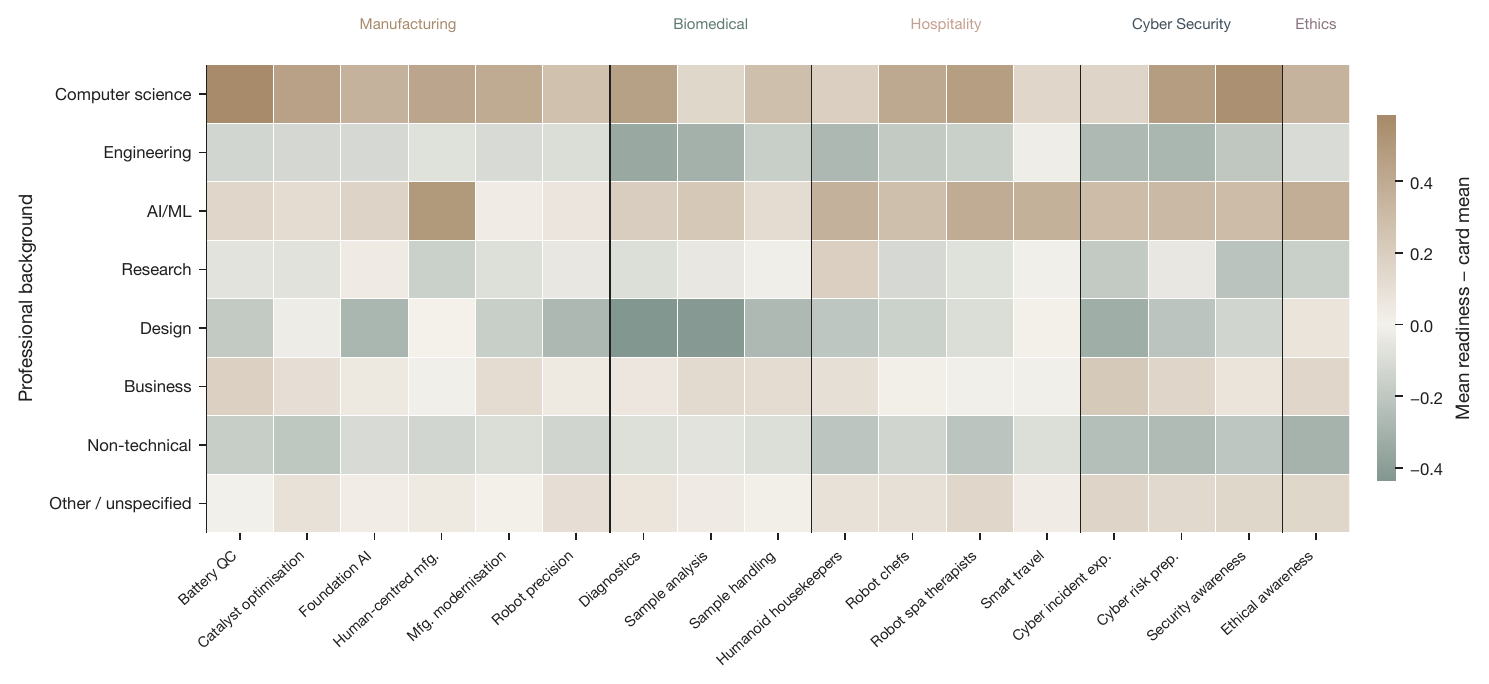}
\caption{Unadjusted mean readiness deviations from each challenge mean, by professional background. Warm: above challenge average; cool: below. Challenges are grouped by challenge family.}
\label{fig:heat}
\end{figure}

Figure~\ref{fig:bgsec} shows mean readiness by reported professional background across the five challenge families. Similar polygon \emph{shape} with different \emph{size} is the visual counterpart of a general background gradient: computer-science and AI/ML respondents sit outside non-technical respondents on every axis, rather than only in some families. The inner polygon is the policy-relevant shape. Non-technical respondents are not drawing a different map of the challenge families; they are giving a more cautious reading of community preparedness on every axis. Because readiness here is perceived community resources and support, including educational resources, not personal refusal to use a technology, that caution is better read as a capability and literacy signal than as technophobia. Figure~\ref{fig:heat} maps background-by-challenge deviations from each challenge mean. These are unadjusted descriptive displays: background is self-reported, groups are unequal, and differences may also reflect age, experience, education or other unmodelled characteristics.

Table~\ref{tab:bgsec} reports clustered contrasts from person-family means (982 respondents; 4{,}532 observed person-family records), comparing each background with non-technical respondents within each challenge family. Computer-science respondents have higher estimated means in every family (contrasts $0.38$-$0.65$, all $p<.01$; all remain $p_{\mathrm{BH}}<.05$). AI/ML respondents have higher estimated means in Manufacturing, Hospitality, Cyber Security and Ethics, with a weaker Biomedical contrast. Business respondents differ most in Cyber Security and Ethics. Engineering and research contrasts are small and not distinguishable from zero. The Manufacturing engineering contrast is $0.02$ ($p=0.86$): in this sample, engineering backgrounds do not speak for a more ready manufacturing community than non-technical respondents do. A manufacturing-readiness exercise whose panel is heavy in computer science or AI/ML will nevertheless read more ready (Manufacturing contrasts $0.55$ and $0.31$) without that shift coming from engineering voices. Twelve of thirteen contrasts with unadjusted $p<.05$ remain below $0.05$ after Benjamini-Hochberg adjustment. The joint Wald test does not reject equality of the background contrasts across challenge families ($p=0.36$). This is absence of evidence for background-by-family heterogeneity, not proof that the contrasts are identical. The estimates support a descriptive cross-family background gradient more clearly than a family-specific pattern, including in Manufacturing. Read as a policy signal rather than as a personality trait, the inner non-technical polygon is a reason to attach literacy and skills programmes to readiness reporting, so that groups who already judge community preparedness more cautiously are not left outside the technology dividend.

\begin{table}[htbp]
\centering
\caption{Associational background contrasts in person-family mean readiness versus non-technical respondents (OLS with respondent-clustered SEs). The model adjusts for challenge family but not for other demographic differences. Manufacturing and non-technical are the reference cells for the underlying interaction model; reported estimates are family-specific total contrasts. $p_{\mathrm{BH}}$ is Benjamini-Hochberg adjusted across the full contrast grid. Joint test of background$\times$family interactions: $p=0.36$.}
\label{tab:bgsec}
\scriptsize
\rowcolors{2}{white}{rowblue}
\begin{tabular}{llrrrrr}
\toprule
Background vs non-technical & Sector & $n$ & Contrast & SE & $p$ & $p_{\mathrm{BH}}$ \\
\midrule
Engineering & Manufacturing & 45 & 0.021 & 0.119 & 0.860 & 0.919 \\
Engineering & Biomedical & 44 & -0.189 & 0.176 & 0.284 & 0.417 \\
Engineering & Hospitality & 47 & 0.004 & 0.136 & 0.976 & 0.976 \\
Engineering & Cyber Security & 53 & 0.023 & 0.155 & 0.883 & 0.919 \\
Engineering & Ethics & 46 & 0.194 & 0.196 & 0.324 & 0.450 \\
Computer science & Manufacturing & 46 & 0.545 & 0.121 & <.001 & <.001 \\
Computer science & Biomedical & 46 & 0.384 & 0.147 & 0.009 & 0.020 \\
Computer science & Hospitality & 47 & 0.470 & 0.145 & 0.001 & 0.004 \\
Computer science & Cyber Security & 49 & 0.639 & 0.145 & <.001 & <.001 \\
Computer science & Ethics & 46 & 0.650 & 0.213 & 0.002 & 0.006 \\
AI/ML & Manufacturing & 36 & 0.314 & 0.122 & 0.010 & 0.021 \\
AI/ML & Biomedical & 35 & 0.274 & 0.168 & 0.104 & 0.173 \\
AI/ML & Hospitality & 36 & 0.516 & 0.150 & <.001 & 0.002 \\
AI/ML & Cyber Security & 38 & 0.579 & 0.153 & <.001 & <.001 \\
AI/ML & Ethics & 36 & 0.675 & 0.178 & <.001 & <.001 \\
Research & Manufacturing & 70 & 0.075 & 0.100 & 0.451 & 0.563 \\
Research & Biomedical & 69 & 0.034 & 0.130 & 0.796 & 0.904 \\
Research & Hospitality & 72 & 0.161 & 0.106 & 0.128 & 0.200 \\
Research & Cyber Security & 76 & 0.065 & 0.116 & 0.573 & 0.682 \\
Research & Ethics & 71 & 0.132 & 0.160 & 0.409 & 0.538 \\
Business & Manufacturing & 150 & 0.227 & 0.078 & 0.004 & 0.010 \\
Business & Biomedical & 151 & 0.199 & 0.102 & 0.052 & 0.093 \\
Business & Hospitality & 155 & 0.191 & 0.093 & 0.040 & 0.076 \\
Business & Cyber Security & 163 & 0.387 & 0.092 & <.001 & <.001 \\
Business & Ethics & 151 & 0.451 & 0.109 & <.001 & <.001 \\
\bottomrule
\end{tabular}
\rowcolors{2}{white}{white}

\end{table}

\subsection{Exploratory card network and comment vocabularies}

\begin{figure}[htbp]
\centering
\includegraphics[width=0.96\textwidth,height=0.42\textheight,keepaspectratio]{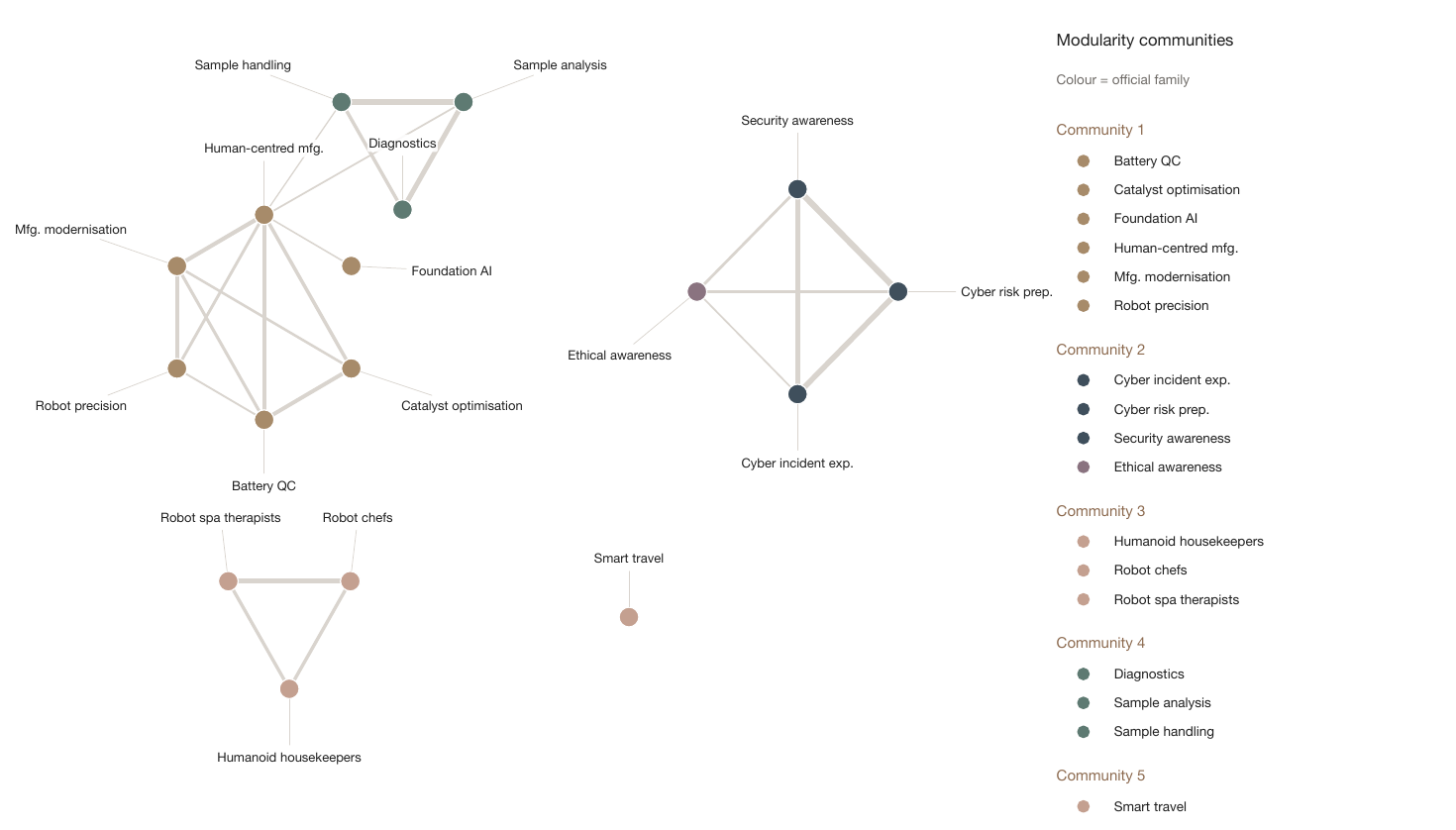}
\caption{Exploratory between-person Spearman network of the 17 challenges (displayed edges: $\rho\geq0.40$). Node colour denotes the instrument challenge family and groups are identified by greedy modularity. Correlations include general response tendencies, and the threshold is a display choice rather than an inferential cutoff.}
\label{fig:net}
\end{figure}

Figure~\ref{fig:net} and Table~\ref{tab:net} show the exploratory correlation network of the 17 challenges. At the prespecified display threshold ($\rho\geq0.40$), the modularity communities largely align with the instrument challenge families, with two exceptions: Ethical Awareness joins the three Cyber Security challenges, and Smart travel is detached from the other Hospitality applications (housekeepers, chefs and spa therapists). Manufacturing cards remain internally grouped and are connected to Biomedical, consistent with their challenge-family profile correlation ($\rho\approx 0.57$). Inside that Manufacturing clump, however, Foundation AI Decision Support is the weakest robotics correlate, matching the family-profile result that a sector-neutral decision-support card should not be treated as a shop-floor indicator. These are features of this thresholded sample network, not a recovered population taxonomy. The Ethics-Cyber grouping is present at thresholds from $0.30$ to $0.45$ and in 100 of 100 person-bootstrap samples at $0.40$; it is not present at $0.50$ (Supplementary Table~S3).

\begin{table}[htbp]
\centering
\caption{Modularity communities in the thresholded card-correlation network. At $\rho\geq0.40$, Ethical Awareness joins Cyber Security and Smart travel separates from the other Hospitality cards.}
\label{tab:net}
\scriptsize
\rowcolors{2}{white}{rowblue}
\begin{tabular}{cll}
\toprule
Community & Official family & Challenge \\
\midrule
1 & Manufacturing & Battery QC \\
1 & Manufacturing & Catalyst optimisation \\
1 & Manufacturing & Foundation AI support \\
1 & Manufacturing & Human-centred manufacturing \\
1 & Manufacturing & Manufacturing modernisation \\
1 & Manufacturing & Robot precision \\
2 & Cyber Security & Cyber incident experience \\
2 & Cyber Security & Cyber risk preparedness \\
2 & Cyber Security & Security awareness \\
2 & Ethics & Ethical awareness \\
3 & Hospitality & Humanoid housekeepers \\
3 & Hospitality & Robot chefs \\
3 & Hospitality & Robot spa therapists \\
4 & Biomedical & Diagnostics \\
4 & Biomedical & Sample analysis \\
4 & Biomedical & Sample handling \\
5 & Hospitality & Smart travel \\
\bottomrule
\end{tabular}
\rowcolors{2}{white}{white}

\end{table}

The three Cyber Security challenges correlate at $\rho=0.59$-$0.63$ (Table~\ref{tab:cyberw}). Ethics cannot be assessed within family because it is represented by one challenge. Its raw correlations with the three Cyber Security challenges are $\rho=0.42$-$0.47$, which places it in the Cyber community at the chosen threshold. These raw between-person associations can partly reflect a respondent's general tendency to give higher or lower readiness ratings; they should not be interpreted as evidence that ethics and cyber security are a single latent construct.

\begin{table}[htbp]
\centering
\caption{Between-person Spearman correlations of readiness among the Cyber Security challenges.}
\label{tab:cyberw}
\small
\rowcolors{2}{white}{rowblue}
\begin{tabular}{llr}
\toprule
Card A & Card B & Spearman $\rho$ \\
\midrule
Cyber risk preparedness & Security awareness & 0.63 \\
Cyber incident experience & Cyber risk preparedness & 0.60 \\
Cyber incident experience & Security awareness & 0.59 \\
\bottomrule
\end{tabular}
\rowcolors{2}{white}{white}

\end{table}

\begin{figure}[htbp]
\centering
\includegraphics[width=0.62\textwidth,height=0.36\textheight,keepaspectratio]{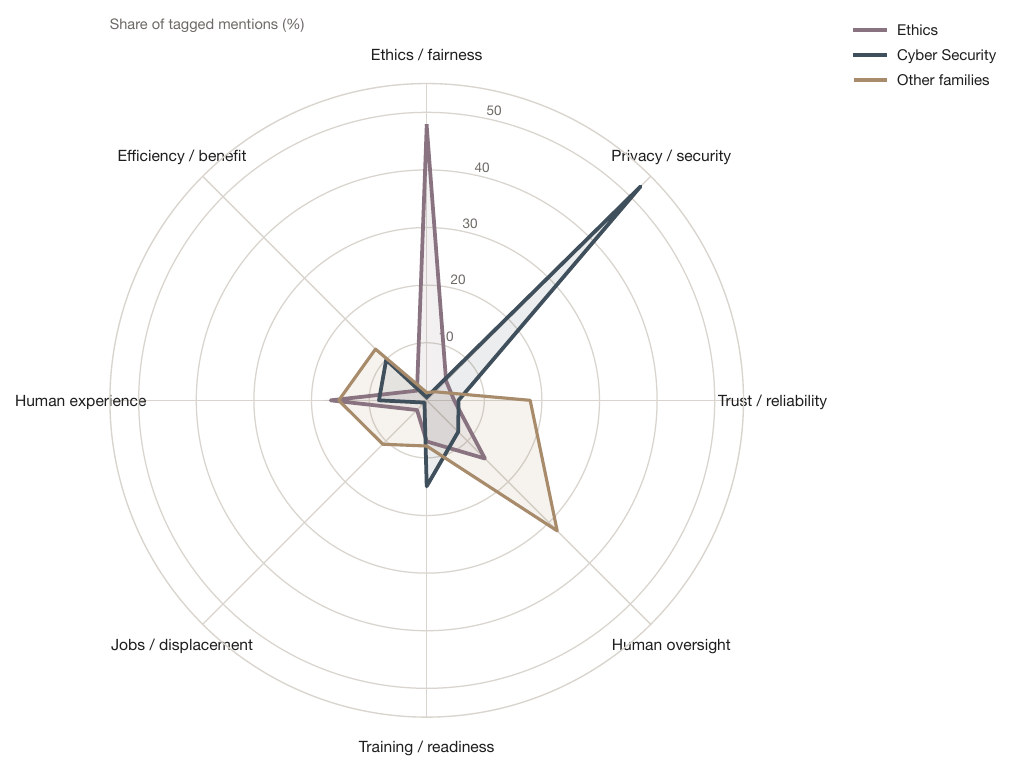}
\caption{Keyword-theme mix in optional free-text comments (Ethics $n=66$; Cyber Security $n=248$; other families $n=803$). Axes are shares of tagged mentions. Tags may overlap, and percentages describe the observed comments rather than a representative qualitative sample.}
\label{fig:ecft}
\end{figure}

The optional comments do not reproduce a common Ethics-Cyber vocabulary (Figure~\ref{fig:ecft}; $n=1{,}117$ observed comments). Among the keyword-tagged mentions, Ethics comments emphasise fairness, bias and accountability (47.6\%; $n=66$ comments), whereas Cyber Security comments emphasise privacy and security (52.5\%; $n=248$). Comments on the other families more often receive human-oversight (32.0\%) and trust (18.0\%) tags. Because comments are optional and the lexicon is coarse, this is descriptive triangulation only. It suggests that a positive co-rating pattern need not imply that respondents frame the two areas in the same terms.

\section{Discussion}

The main finding is not simply that averages can conceal disagreement. It is that differences among broad challenge-family means account for only a small part of the variation in perceived AI and robotics Readiness observed in this survey. Respondents frequently disagree about the same named challenge, and the crossed decomposition shows that stable differences among respondents account for substantially more variation than differences among the 17 challenge means. The family-level decomposition sharpens the point further: differences among challenge-family means account for only about 2\% of observed variation, while substantially more variation occurs through who is judging, how individuals judge a family, and how particular applications are evaluated within it. A sector or domain ranking may therefore be useful for portfolio scanning, but it is not a summary of most of the information contained in the survey.

This distinction matters directly for policy and organisational decision making. A high mean Readiness score can describe very different situations. It may reflect relatively broad agreement, or it may average over respondents who see the same application very differently. Likewise, two applications grouped under the same sector label may not represent the same preparedness problem. Readiness assessments should therefore report both the level and the dispersion of ratings and retain application-level evidence where funding, workforce development, deployment or governance decisions concern a specific technology. The appropriate unit of analysis should follow the decision being made: sector summaries may be adequate for portfolio scanning, while application decisions require application-level evidence.

\subsection{A high sector score is not necessarily a shared preparedness judgement}

Manufacturing provides the clearest example because it is the application domain most likely to appear safely aggregable. It has the highest mean and lower within-person application dispersion than Hospitality, but the underlying applications remain distinct. The shop floor and industrial-robotics cards form a relatively coherent core, while catalyst-process optimisation is rated less ready and Foundation AI Decision Support has the highest mean but is only loosely associated with the robotics applications. A single Manufacturing score therefore risks obscuring where support or investment may actually be needed.

Hospitality shows the same problem more visibly. Smart Travel receives a substantially different profile from the service-robot applications, while Humanoid Housekeepers and Robot Chefs are among the most dispersed challenges in the survey. These results do not imply that sector statistics are useless. They show that a sector label should not automatically be treated as the unit of preparedness. For policymakers and sector bodies, the practical question is whether the applications grouped under a label are sufficiently similar for the intended decision. For organisations, readiness exercises should therefore retain application-level results whenever deployment, training or investment choices concern particular systems rather than an entire portfolio.

\subsection{Whose ratings form the average also matters}

Professional composition changes the level of reported Readiness. Computer science and AI/ML respondents give higher assessments than non-technical respondents across several challenge families, while engineering respondents do not report higher Manufacturing Readiness than non-technical respondents. The joint background-by-family interaction is not significant, suggesting a broad professional-background gradient rather than a completely different professional map for each challenge family.

This has a straightforward implication for consultation and readiness exercises. A panel dominated by digitally specialised respondents can produce a higher readiness mean even when engineering and non-technical respondents do not share that assessment. Reports should therefore disclose who contributed to the score and show important group-specific estimates rather than presenting the average as a common judgement. For policymakers, more cautious non-technical assessments should not automatically be interpreted as resistance to technology: because Readiness concerns perceived community resources and support, they may instead signal unmet literacy, skills, access or capability needs. For organisations, workforce development should therefore accompany readiness measurement rather than being designed only around the groups already most confident in the technology environment.

\subsection{Ethics and Cyber Security should inform decisions without being collapsed}

Ethical Awareness and Cyber Security move together in the observed ratings, but much of the raw cross-domain correlation disappears after accounting for respondents' general tendency to give higher or lower ratings. The remaining Ethics-Cyber association is positive but modest ($r=0.11$). Both are also positively associated with application-domain Readiness, but there is no clear evidence of a strong Ethics$\times$Cyber interaction or of substantially different joint relationships across Biomedical, Manufacturing and Hospitality.

The practical implication is that ethics and cyber security should enter AI and robotics decisions, but they should not be treated as one preparedness construct. The free-text evidence supports this distinction: ethics comments emphasise fairness, bias and accountability, whereas cyber comments focus primarily on privacy and security. Policymakers and organisations should therefore consider both when assessing deployment conditions, while diagnosing their underlying concerns separately. Numerical co-movement does not imply that the same intervention addresses both.

\subsection{From readiness scores to decision diagnostics}

Taken together, the results suggest a different use for readiness assessment. A readiness score should be treated as the start of diagnosis rather than the end of assessment. Decision makers should ask four questions: how high is the average, how much disagreement surrounds it, which applications are being combined, and whose judgements form the score. These questions determine whether a sector-level summary is sufficient or whether the evidence should be disaggregated before investment, training, regulation or deployment decisions are made.

For policymakers, this means avoiding league-table interpretations in which a higher sector mean automatically implies that an industry requires less support. A high mean with substantial disagreement may still indicate uneven capability, contested implementation conditions or excluded groups. For sector bodies, it means checking whether applications grouped under the same heading are actually judged similarly before communicating one sector score. For organisations, it means matching the reporting unit to the implementation decision and showing the composition of the respondent panel. The purpose of disaggregation is not to eliminate summary indicators, but to prevent a convenient summary from being mistaken for consensus.

\subsection{Limitations}

The 17 challenges are a purposive rather than representative sample of all AI and robotics applications, and the challenge families contain unequal numbers and types of cards. Ethical Awareness is represented by a single card and the fixed card order creates possible order and fatigue effects. Significance, Complexity and Readiness are single-item measures presented in the same exercise, creating common-method risk.

The crossed residual combines person-by-challenge variation with measurement error because there is only one rating per person-challenge cell, while the challenge-family variance partition is an approximate descriptive decomposition. The correlation network and optional free-text analysis are exploratory, and professional-background contrasts are associational and not fully adjusted for all possible demographic differences. The Ethics-Cyber analyses likewise identify associations rather than causal pathways. These limitations affect how precisely the sources of disagreement can be identified, but they do not alter the central descriptive finding that broad challenge-family means capture only a small part of the variation observed in this readiness exercise.

\section{Conclusion}

AI and robotics Readiness scores at the sector or domain level can be useful summaries, but broad challenge-family differences capture only a small part of the variation in this survey. Differences among challenge-family means account for about 2\% of observed variation, while substantially more variation lies in who is judging, how individuals judge a family, and how particular applications are evaluated. Even Manufacturing, the highest-rated and comparatively more coherent application domain, combines applications that should not automatically be treated as one preparedness object.

The implication for decision makers is simple. Policymakers should use sector scores to scan portfolios, not to declare industries uniformly ready or behind. Sector bodies should check whether applications grouped under one label are judged similarly before communicating a single readiness number. Organisations should report who contributed to the assessment and retain application-level disagreement when making deployment, workforce or investment decisions. Ethics and Cyber Security should also inform those decisions, but should not be collapsed into one concern.

The key question is therefore not simply \emph{which sector is most ready?} It is \emph{ready for which application, according to whom, and with how much agreement?}

\section*{Funding}
This work was supported by the Northwest Crucible Pump Priming scheme, delivered by the Horizons Institute. The funder had no role in study design; data collection, analysis or interpretation; preparation of the manuscript; or the decision to submit the article for publication.

\section*{Acknowledgements}
The author thanks Naomi Adel, Amy E.\ Morgan, Folayo Aina, Demos Parapanos, Vikas Mackevicius and Teslim Olayiwola Salahudeen for their contributions to the shared survey design and data collection that this paper builds on. The author also acknowledges the Northwest Crucible Pump Priming scheme and the Horizons Institute for supporting this research. The views expressed are those of the author and do not necessarily reflect those of the funder.

\section*{Declaration of competing interests}
The author declares that he has no known competing financial interests or personal relationships that could have appeared to influence the work reported in this paper.

\section*{Data availability}
De-identified analysis files, card instruments and analysis code will be deposited in a public repository upon acceptance, subject to the consent and ethics terms of the study. Until deposit, they are available from the corresponding author. The Supplementary Material is included at the end of this preprint.

\section*{Ethics}
Participants were recruited through Prolific between 12 and 29 June 2026, provided informed consent and were compensated according to the platform listing. Responses were analysed in de-identified form.

\bibliography{references,references_additions,references_tis_review_additions}

\clearpage
\begin{center}
{\Large\bfseries Supplementary Material}
\end{center}

\noindent Additional tables and instrument notes for: Whose readiness counts? Disagreement within and between sectors in perceived AI and robotics preparedness.

\section*{S1. Challenge titles and core items}

The 17 challenges, with instrument families, are: Security awareness, Cyber risk preparedness, Cyber incident experience (Cyber Security); Ethical awareness (Ethical Awareness); Diagnostics, Sample handling, Sample analysis (Biomedical); Smart travel, Humanoid housekeepers, Robot chefs, Robot spa therapists (Hospitality); Catalyst optimisation, Battery QC, Human-centred manufacturing, Foundation AI Decision Support, Manufacturing modernisation, Robot precision (Manufacturing). Card order was fixed. Significance, complexity and readiness appeared together on each card using the five-point anchors given in the main text. Optional comments were not required.

\section*{S2. Within-family mean pairwise absolute differences}

Mean absolute difference of readiness between every pair of applications in a multi-card family, averaged across pairs and across respondents with both ratings. Pairwise $n$ is at least 866 in every family.

\rowcolors{2}{white}{rowblue}
\begin{tabular}{lrrr}
\toprule
Family & Cards & Mean pairwise $|R_a-R_b|$ & Min pair $n$ \\
\midrule
Hospitality & 4 & 1.07 & 898 \\
Manufacturing & 6 & 0.76 & 868 \\
Biomedical & 3 & 0.70 & 866 \\
Cyber Security & 3 & 0.64 & 916 \\
\bottomrule
\end{tabular}
\rowcolors{2}{white}{white}

\section*{S3. Network threshold sensitivity}

Greedy modularity communities in the between-person Spearman network at alternative display thresholds. ``Ethics with Cyber'' means Ethical awareness shares a community with at least one Cyber Security challenge. At $\rho=0.40$, Ethical awareness shared a community with all three Cyber Security challenges in 100 of 100 person-bootstrap samples.

\rowcolors{2}{white}{rowblue}
\begin{tabular}{lrrr}
\toprule
Threshold $\rho$ & Edges & Communities & Ethics with Cyber \\
\midrule
0.30 & 84 & 2 & yes \\
0.35 & 50 & 3 & yes \\
0.40 & 24 & 5 & yes \\
0.45 & 19 & 6 & yes \\
0.50 & 13 & 7 & no \\
\bottomrule
\end{tabular}
\rowcolors{2}{white}{white}

\end{document}